\documentclass{webofc}
\usepackage[varg]{txfonts}   

\usepackage{dsfont}

\woctitle{CNR*15}
\begin{document}
\title{ Fission life-time calculation using a complex absorbing potential }
%
%

\author{Guillaume Scamps \inst{1}\fnsep\thanks{\email{scamps@nucl.phys.tohoku.ac.jp }} \and
        Kouichi Hagino \inst{1,2}\fnsep\thanks{\email{hagino@nucl.phys.tohoku.ac.jp}} 
}

\institute{Department of Physics, Tohoku University, Sendai 980-8578, Japan
\and
           Research Center for Electron Photon Science, Tohoku University, 1-2-1 Mikamine, Sendai 982-0826, Japan 
          }

\abstract{%

A comparison between the semi-classical approximation and the full quantum calculation with a complex absorbing potential is made with a model of the fission of $^{258}$Fm.
The potential barrier is obtained with the constrained Skyrme HF+BCS theory. The life-time obtained by the two calculations agree with each other the difference being only by 25\%.

}
\maketitle

\section{Introduction}
\label{intro}

The description of the quantum tunneling is important to predict the life-time of a spontaneous fission as well as fission product yields.
A standard strategy to compute the fission life-time is to first determine a potential energy surface using the liquid drop model with shell correction \cite{Brack72,Moller01,Moller09}  or the self-consistent mean field theory \cite{Skalski08,Staszcak09,Baran11,Sadhukhan13,Mcdonnell14,Sad14,Warda12,Robledo14,Giuliani14,Goutte05,Goutte11,Afanasjev10,Lu12,Zhao15}. 
The life-time is then calculated using the semi-classical approximation to determine the fission path that minimizes the action \cite{Brack72, Sadhukhan13,Sad14,Sch86,KI89,I94}. 

In a previous study \cite{sca15}, we developed a method to obtain the fission life-time using a fully quantum approach with a complex absorbing potential. This method has been tested on a schematic model.

In the present work, we test the semi-classical approximation in a more realistic application, by comparing the full quantum approach to the semi-classical calculation using a potential determined with the constrained Hartree-Fock method with a Skyrme interaction.

\section{Complex absorbing potential method}
\label{sec-1}

In order to describe the evolution of a quantum system into the continuum spectrum, a possibility is to add an imaginary potential at larges distances in order to absorb the wave function, in which the absorbed component of the wave function is assumed to be emitted in the continuum. The modified hamiltonian $\hat H'= \hat H+iW(R)$ is then composed of the original hamiltonian $\hat H$ and the imaginary potential $iW(R)$. The time dependent Shr\"odinger equation for the system becomes,
\begin{align}
i \hbar \frac{d}{dt} |\Psi(t) \rangle = \left( \hat{H}+iW(R) \right) |\Psi(t) \rangle.
\end{align}
This time-dependent equation is often solved numerically with the Runge-Kutta or other iterative methods. Nevertheless, those iterative methods are limited to the case where the phenomenon of interest takes place in a time scale of the order of magnitude of the characteristic time scale of the system. This is not the case for the spontaneous fission, where the fission life time can be of the order magnitude of years while  the characteristic time scale of the nuclear system is the zepto second .

Alternatively,  one can integrate the time dependent equation as,
\begin{align}
	|\Psi(t) \rangle = e^{-\frac{i}{\hbar}t \hat{H}'} | \Psi_0 \rangle.
\label{eq:propagator}
\end{align}
In order to compute easily the exponential of the matrix, we expand the hamiltonian $H'$  on the biothogonal basis \cite{Mor52,Hussein95,HT98} with the left and right eigenfunctions,
\begin{align}
	H' | \varphi^r_i \rangle = E_i  | \varphi^r_i \rangle \quad {\rm and} \quad   \langle \varphi^l_i | H' = E_i  \langle  \varphi^l_i | , \label{eq:eig_left_right}
\end{align}
with complex eigenvalue $E_i$. Using the completeness relation for the bi-orthogonal basis $\sum_i| \varphi^r_i \rangle \langle  \varphi^l_i | = \mathds{1} $, we obtain the simple evolution,
\begin{align}
	| \Psi(t) \rangle = \sum_i e^{-\frac{i}{\hbar}t E_i}  \langle  \varphi^l_i  | \Psi_0 \rangle   
| \varphi^r_i \rangle \label{eq:TD_diag} .
\end{align}
It is then possible to describe easily the evolution of the wave packet for a very long time once we compute the overlap $\langle  \varphi^l_i  | \Psi_0 \rangle $.

We note that each eigenstate $| \varphi^r_i \rangle$ is associated with a complex energy, $E_i=E_i^r-i\Gamma_i/2 $, with $\Gamma$ being the resonance width which is related to the life time, $\tau_i$=$\hbar / \Gamma_i$.

\section{Results}

In the precedent study \cite{sca15}, we tested this method in a model case consisting of a schematic barrier. In the present study, we use a more  realistic potential using the constrained Hartree-Fock+BCS theory.
The calculation is carried out for the $^{258}$Fm with the Skyrme interaction Skm* and a volume type of pairing. The  constrained Hartree-Fock BCS  equation is solved using the  {\rm EV8} code \cite{Bon05}. A constraint is applied on the value of the average separation $R$ of the two fragments as in Ref. \cite{sim14} assuming a symmetric fission with a neck position  at $z$=0.

\begin{figure}
\centering
\includegraphics[width=7cm,clip]{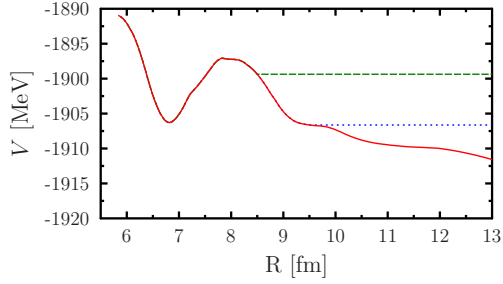}
\caption{  The potential energy as a function of the separation between the two fragments for the symmetric fission of $^{258}$Fm. The original potential is shown with the red  solid line. The modified potential for the initial wave function and dynamics are shown respectively by the green dashed line and the blue dotted line. }
\label{fig:pot}       
\end{figure}

The obtained potential is shown in fig. \ref{fig:pot}. The collective hamiltonian is assumed here to have only one dimension,
\begin{align}
H(R) &= - \frac{\hbar^2}{2M} \frac{\partial^2}{\partial R^2} + V(R).
\end{align}
In this equation the mass $M$ is assumed to be a constant with a value to reproduce the experimental spontaneous half-life of $^{258}$Fm, $\tau_{\rm exp}$=0.37 ms \cite{Hol00}, using the semi classical approximation. With this approximation, the life time is determine as,
\begin{align}
  \tau = \frac{2 \pi}{\Omega}\,e^{-2 S/\hbar},
\label{eq:WKB_1D}
\end{align}
with $\Omega=2(E_0-V_0)/\hbar$  and the action integral $S$,
\begin{align}
  S=\int^{R_1}_{R_0}\sqrt{2 M (U(R)-E_0)}\,dR,
\end{align}
Here, $R_0$ and $R_1$ are the inner and outer turning points, respectively defined as $V(R_0)=V(R_1)=E_0$, $E_0$ is  taken as the energy obtained by solving the static Schr\"odinger  equation by restricting the space before the barrier.  The mass that reproduces the experimental life-time is $M$= 480 MeV\,(10$^{-22} s)^2$\,fm$^{-2}$. As mentioned earlier, the goal of the present contribution is not to predict the fission life time but to show the feasibility of this approach and to compare the result with   the semi-classical approximation.

\begin{figure}
\centering
\includegraphics[width=7cm,clip]{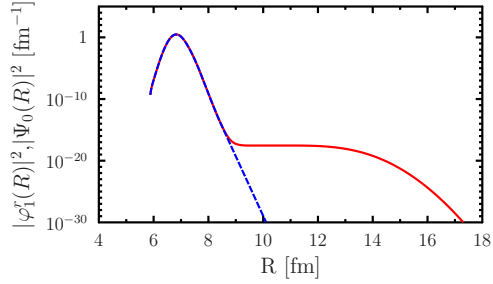}
\caption{  The initial wave function (the blue dashed line) and the resonant wave function (the red solid line).  }
\label{fig:waves}       
\end{figure}

 To solve the tunneling problem, we first determine the initial wave function $|\Psi_0\rangle$ with a modified potential in order to have a bound state inside the barrier. That is the potential is replaced after the barrier by a constant value outside the barrier, after $R=$8.5 fm.
The lowest energy bound state is chosen as the initial wave function and is shown in fig. \ref{fig:waves}.

To obtain the resonant state, an imaginary potential is then added to the hamiltonian \cite{IY05,Ota14},
\begin{align}
  i W(R) = i W_0  (R-R_a)^2\theta(R-R_{a}), \label{eq:potim}
\end{align}
with $W_0$=$-0.1$~MeV.fm$^{-2}$   and $R_a$=$10.8$~fm. We also modified the real potential for the dynamics and is chosen to be constant after $R$=9.5 fm. We take this prescription in order to stabilize the results with the choice of the imaginary potential. That is, with this prescription, the results are independent of the distance between the barrier and the position of the imaginary potential. Among the eigenfunctions of this Hamiltonian, we select the physical resonance wave function
as the  state $|\varphi_1^r\rangle$ which has the maximum overlap with the initial state. From the imaginary part of the energy of this state, we determine the decay half-life. The resultant value is 0.290 ms, which agrees well with the semi-classical approximation, $\tau$=0.37 ms. The difference is only by 25\%

\section{Summary}

A comparison between the semi-classical approximation and the full quantum calculation with a complex absorbing potential was made with a model for the fission of $^{258}$Fm. We found that those two calculations lead to a similar result to each other.
A difference of 25\% was found between the two results, that could be attribute to the assumption of harmonicity for the confining potential or to the accuracy of the semi-classical approximation. 
In this paper, we considered a schematic constant mass for the fission motion, but we plan for a future application to use the mass from a microscopical calculation.

 \section*{Acknowledgement}
 
 G.S. acknowledges the Japan Society for the Promotion of Science
 for the JSPS postdoctoral fellowship for foreign researchers.
 This work was supported by Grant-in-Aid for JSPS Fellows No. 14F04769.

\end{document}